# 基於 NTRU 的後量子密碼學金鑰擴展方法和匿名憑證方案

Abel C. H. Chen, *Senior Member, IEEE*

*摘要*—NTRU 是基於格後量子密碼學的重要方法之一，具備抗量子計算攻擊的能力，並且有加密效率高等優勢。然而，NTRU 的缺點是產製金鑰對的效率較差。有鑑於此，本研究提出基於 NTRU 的金鑰擴展方法，可以具有較高的效率對公鑰進行擴展。除此之外，本研究把基於 NTRU 的金鑰擴展方法應用到匿名憑證方案，讓終端設備僅需產製一次金鑰對，後續可以由憑證中心擴展出多把不同的公鑰，並且其他設備無法從擴展後公鑰推導出原始公鑰，從而保障匿名性，可以應用到對隱私有高度需求的服務。在實驗中，本研究實作和比較多種不同的參數組合，並且證明金鑰擴展效率顯著高於金鑰對產製效率。除此之外，本研究亦和其他密碼學方法比較，證明本研究所提的基於 NTRU 的金鑰擴展方法具有較高的效率。

*關鍵字*—NTRU、後量子密碼學、金鑰擴展方法、匿名憑證方案、蝴蝶金鑰擴展。

## I. 前言

有鑑於量子計算的技術日益成熟，並將對 RSA 密碼學和橢圓曲線密碼學(Elliptic Curve Cryptography, ECC)帶來威脅[1]，所以美國國家標準暨技術研究院(National Institute of Standards and Technology, NIST)開始制定多個後量子密碼學(Post-Quantum Cryptography, PQC)標準[2]。其中，由於基於格(Lattice-based)密碼學方法具有較高的計算效率，所以美國國家標準暨技術研究院陸續制定了基於格後量子密碼學標準，包含基於模格金鑰封裝機制(Module-Lattice-Based Key-Encapsulation Mechanism, ML-KEM)(前身為 CRYSTALS-Kyber)[3]、基於模格數位簽章演算法(Module-Lattice-Based Digital Signature Algorithm, ML-DSA)(前身為 CRYSTALS-Dilithium)[4]、以及正在制定中的基於 NTRU 格快速傅立葉變換數位簽章演算法(FFT (Fast-Fourier Transform) over NTRU-Lattice-Based Digital Signature Algorithm, FN-DSA)(前身為 Falcon)[5]。除此之外，美國國家標準暨技術研究院亦制定了後量子密碼學遷移草案，建議 2030 年起棄用 RSA 密碼學和橢圓曲線密碼學的部分參數[6]。因此，遷移到後量子密碼學是重要的議題，以提升抵抗量子計算攻擊的資訊安全能力。

除了遷移到後量子密碼學之外，在許多應用和服務上還具有隱私的需求，例如：車聯網通訊[7]、金融交易[8]、醫療照護[9]等。因此，IEEE 1609.2.1 標準中設計了蝴蝶金鑰擴展(Butterfly Key Expansion, BKE)方法和假名憑證方案[10]，可以讓終端設備(End Entity, EE)(即車載設備)產製橢圓曲線密碼學金鑰對(稱為毛蟲金鑰對)，把毛蟲公鑰傳送給註冊中心(Registration Authority, RA)，由註冊中心把毛蟲公鑰擴展為繭公鑰，再把繭公鑰傳送給憑證中心(Certificate Authority, CA)，由憑證中心把繭公鑰擴展為蝴蝶公鑰[11]。由於經典電腦(Classical Computer)無法在多項式時間內破解橢圓曲線離散對數問題(Elliptic Curve Discrete Logarithm Problem, ECDLP)[12]，所以從而保障其他設備無法從蝴蝶公鑰反推回毛蟲公鑰，達到匿名性和隱私保護的效果。然而，由於量子計算搭配 Shor 演算法將有機會在多項式時間內破解橢圓曲線離散對數問題[13]，所以設計一個基於後量子密碼學的匿名憑證方案是當前需要突破的研究問題。

有鑑於此，本研究主要在 NTRU 密碼學方法[14]的基礎上，提出基於 NTRU 的後量子密碼學金鑰擴展方法和匿名憑證方案。其中，由於 NTRU 密碼學方法主要的缺點在於產製金鑰對需要花費較多的計算時間，所以本研究可以通過提出的金鑰擴展方法快速產製擴展後公鑰，同時兼具隱私保護且高計算效率的優勢。本研究主要貢獻條列如下：

- 本研究提出基於 NTRU 的後量子密碼學金鑰擴展方法，可以在原始公鑰乘上隨機數得到擴展後公鑰，並且具有比產製金鑰對的效率高數千倍。
- 本研究提出基於 NTRU 的匿名憑證方案，憑證中心可以運用基於 NTRU 的後量子密碼學金鑰擴展方法對原始公鑰進行擴展，並且在匿名憑證中的公鑰欄位放置擴展後公鑰，可以避免被其他設備從擴展後公鑰反推原始公鑰，進而達到匿名效果。
- 實驗比較本研究提出基於 NTRU 的後量子密碼學金鑰擴展方法、基於橢圓曲線密碼學的金鑰擴展方法[10]、基於編碼(Code-based)的後量子密碼學金鑰擴展方法[15]，並且由實驗結果顯示本研究所提方法具有較高的效率。

本文主要分為五節。第 II 節介紹 NTRU 密碼學方法，第 III 節提出基於 NTRU 的後量子密碼學金鑰擴展方法和匿名憑證方案，並且詳細說明其原理。第 IV 節進行實驗證明，通過實證研究來比較不同演算法的效能。最後，第 V 節總結本研究發現和討論未來研究方向。

Abel C. H. Chen is with the Information & Communications Security Laboratory, Chunghwa Telecom Laboratories, Taoyuan 326, Taiwan *(Corresponding author: Abel C. H. Chen).*

## II. NTRU 密碼學方法

本研究將於第 II.A 節介紹金鑰對產製方法，第 II.B 節介紹加密方法和解密方法。其中，在 NTRU 密碼學[14]將會先選定參數整數 $p$ 和整數 $q$，並且最大公因數為 1 (即 $GCD(p, q) = 1$)，以及多項式的階 $N$，然後產製 3 個多項式環，分別為 $R = \frac{\mathbb{Z}[x]}{(x^N-1)}$、$R_p = \frac{\mathbb{Z}_p[x]}{(x^N-1)}$、$R_q = \frac{\mathbb{Z}_q[x]}{(x^N-1)}$，在此基礎上進行計算。

### A. 金鑰對產製方法

在產製金鑰對的過程中，將先產製一個隨機多項式 $f(x) \in R$，再基於多項式$f(x)$產製多項式$F_p(x)$和多項式$F_q(x)$，需符合公式(1)和(2)定義，並且把$\{f(x), F_p(x)\}$作為私鑰。為了產製公鑰，將先產製另一個隨機多項式 $g(x) \in R$，並且與多項式$F_q(x)$相乘後得到多項式作為公鑰$h(x)$，如公式(3)所示。其中，從$h(x)$反推$F_q(x)$將可以歸納為最小向量問題(Shortest Vector Problem, SVP)[14]，並且目前尚無方法可以在多項式時間解決最小向量問題，從而保障攻擊者無法從公鑰推導出私鑰。

$$F_p(x) \in R_p \equiv f(x)^{-1} \pmod{p}(\mod x^N - 1) \quad (1)$$
$$F_q(x) \in R_q \equiv f(x)^{-1} \pmod{q}(\mod x^N - 1) \quad (2)$$
$$h(x) \in R_q \equiv F_q(x)g(x) \pmod{q}(\mod x^N - 1) \quad (3)$$

### B. 加密方法和解密方法

在加密的過程中，為了達到選擇明文攻擊下的不可區分性(INDistinguishability under Chosen Plaintext Attack, IND-CPA)，在每次加密時都會加入隨機數，讓同一公鑰對同一明文加密後，每次的密文可以不同。因此，首先將產製隨機多項式$b(x) \in R_q$，然後把明文解碼為多項式的係數值產製待加密訊息多項式$M(x) \in R_p$。透過公式(4)，可以運用公鑰$h(x)$對待加密訊息多項式$M(x)$進行加密，產製多項式$C(x)$作為密文[14]。後續如果有需要儲存和傳送時，可以傳送多項式$C(x)$的係數值。其中，由於每次產製的隨機多項式$b(x)$不同，所以可以提升密文的安全性。

$$\begin{aligned}C(x) &\in R_q \\ &\equiv p \times h(x) \times b(x) + M(x) \pmod{q}(\mod x^N - 1) \\ &\equiv p \times F_q(x)g(x) \times b(x) + M(x) \\ &\quad \pmod{q}(\mod x^N - 1)\end{aligned} \quad (4)$$

在解密的過程中，為了消除隨機多項式$b(x)$和保留待加密訊息多項式$M(x)$，所以通過公式(5)、公式(6)、公式(7)計算[14]。其中，公式(5)主要將密文$c(x)$和私鑰$f(x)$相乘後模 $q$，然後公式(6)可以把公式(5)結果轉換到$R_p$環的多項式$\tau(x)$，然後公式(7)可以把公式(6)結果$\tau(x)$和私鑰$F_p(x)$相乘後模 $p$，即可還原得到待加密訊息多項式$m(x)$。

$$\begin{aligned}C(x) \times f(x) &\in R_q \\ &\equiv p \times F_q(x)g(x) \times b(x) \times f(x) + M(x) \times f(x) \\ &\quad \pmod{q}(\mod x^N - 1) \quad (5)\\ &\equiv p \times g(x) \times b(x) + M(x) \times f(x) \\ &\quad \pmod{q}(\mod x^N - 1)\end{aligned}$$

$$\tau(x) \in R_p = C(x) \times f(x) \pmod{p}(\mod x^N - 1) \quad (6)$$

$$\begin{aligned}\tau(x) \times F_p(x) &\in R_p \\ &\equiv p \times g(x) \times b(x) \times F_p(x) + M(x) \times f(x) \times F_p(x) \\ &\quad \pmod{p}(\mod x^N - 1) \quad (7)\\ &\equiv M(x) \in R_p\end{aligned}$$

## III. 本研究提出的方法

本研究將先於第 III.A 節提出基於 NTRU 的後量子密碼學金鑰擴展方法，說明金鑰擴展方法的核心精神。之後，第 II.B 節再提出完整的基於 NTRU 的匿名憑證方案，可以達到對憑證中心匿名。

### A. 基於NTRU 的後量子密碼學金鑰擴展方法

#### 1) 方法流程及其原理

本研究提出的基於 NTRU 的後量子密碼學金鑰擴展方法(如圖 1)，其使用參數如第 II 節所定義。首先，將先由終端設備產製原始金鑰對(Original Key Pair)，包含私鑰$\{f(x), F_p(x)\}$和原始公鑰$h(x)$。然後，終端設備再運用憑證中心公鑰加密原始公鑰$h(x)$和相關資訊(如權限(Permissions)) $I$ 得到密文 $c_a$，再把密文 $c_a$ 傳送給憑證中心。

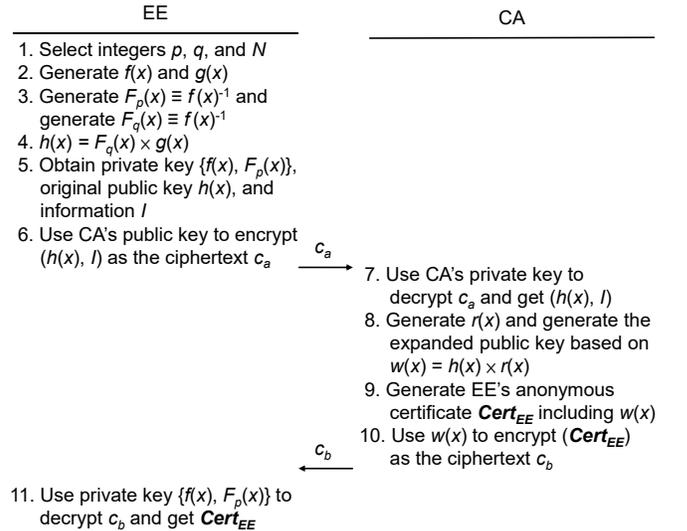

**Fig. 1.** 基於 NTRU 的後量子密碼學金鑰擴展方法

當憑證中心收到密文 $c_a$ 後，用其私鑰解密得到終端設備的原始公鑰$h(x)$和相關資訊 $I$。為了產製擴展後公鑰，將先產製另一個隨機多項式$r(x) \in R$，並且與原始公鑰$h(x)$相乘後得到多項式作為擴展後公鑰$w(x)$，如公式(8)所示。之後，憑證中心產製終端設備的匿名憑證 $Cert_{EE}$，並且在匿名憑證 $Cert_{EE}$ 中的公鑰欄位存放擴展後公鑰$w(x)$，然後再用擴展後公鑰$w(x)$加密匿名憑證 $Cert_{EE}$ 後得到密文 $c_b$，並回傳密文 $c_b$ 給終端設備。

$$w(x) \in R_q \equiv h(x)r(x) \pmod{q}(\mod x^N - 1) \quad (8)$$

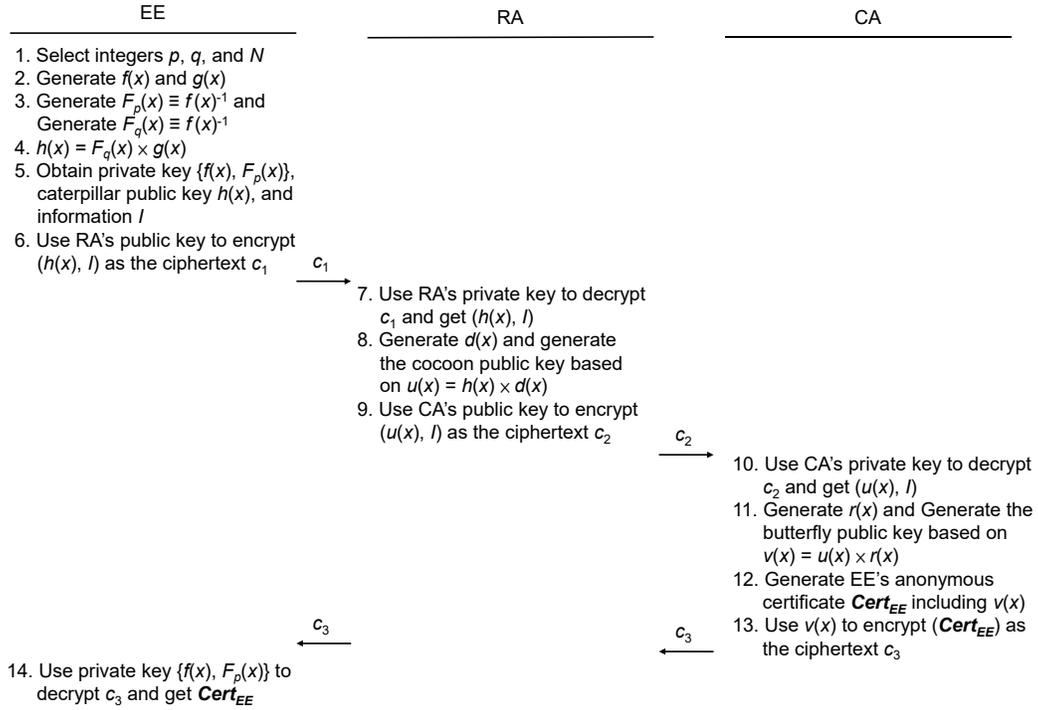

**Fig. 2.** 基於 NTRU 的匿名憑證方案

為說明運用擴展後公鑰$w(x)$加密匿名憑證 $Cert_{EE}$ 的過程中，本節假設匿名憑證 $Cert_{EE}$ 作為明文解碼為多項式的係數值產製待加密訊息多項式$M(x) \in R_p$。透過公式(9)，產製隨機多項式$b(x) \in R_q$，然後運用擴展後公鑰$w(x)$對待加密訊息多項式$M(x)$進行加密，產製多項式$c_b(x)$作為密文。後續主要傳送多項式$c_b(x)$的係數值。

$c_b(x) \in R_q$
$\equiv p \times w(x) \times b(x) + M(x) \pmod{q}\pmod{x^N - 1}$
$\equiv p \times F_q(x)g(x)r(x) \times b(x) + M(x)$
$\pmod{q}\pmod{x^N - 1}$ (9)

當終端設備收到密文 $c_b$ 後，可以運用私鑰$\{f(x), F_p(x)\}$解密得到匿名憑證 $Cert_{EE}$，計算過程如公式(10)、公式(11)、公式(12)。

$c_b(x) \times f(x) \in R_q$
$\equiv p \times F_q(x)g(x)r(x) \times b(x) \times f(x) + M(x) \times f(x)$
$\pmod{q}\pmod{x^N - 1}$ (10)
$\equiv p \times g(x)r(x) \times b(x) + M(x) \times f(x)$
$\pmod{q}\pmod{x^N - 1}$
$\tau(x)' \in R_p = c_b(x) \times f(x) \pmod{p}\pmod{x^N - 1}$ (11)
$\tau(x)' \times F_p(x) \in R_p$
$\equiv p \times g(x)r(x) \times b(x) \times F_p(x) + M(x) \times f(x) \times F_p(x)$
$\pmod{p}\pmod{x^N - 1}$ (12)
$\equiv M(x) \in R_p$

2) 安全性分析

基於 NTRU 的後量子密碼學金鑰擴展方法的核心精神主要建構在擴展後公鑰，如公式(8)所示。由於從$w(x)$反推$h(x)$將可以歸納為最小向量問題[14]，並且目前尚無方法可以在多項式時間解決最小向量問題，從而保障攻擊者無法從擴展後公鑰推導出原始公鑰，達到匿名性。

然而，此方法雖然能做到對憑證中心之外的其他設備匿名，但卻無法對憑證中心匿名，憑證中心可以知道擴展後公鑰$w(x)$和原始公鑰$h(x)$之間的關聯。

*B. 基於NTRU 的匿名憑證方案*

1) 方法流程及其原理

為做到對憑證中心匿名，讓憑證中心無法追蹤出是哪一個終端設備，本研究參考 IEEE 1609.2.1 的蝴蝶金鑰擴展方法[10]提出的基於 NTRU 的匿名憑證方案(如圖 2)，其使用參數如第 II 節和第 III.A 節所定義。首先，將先由終端設備產製毛蟲金鑰對，包含私鑰$\{f(x), F_p(x)\}$和毛蟲公鑰$h(x)$。然後，終端設備再運用註冊中心公鑰加密毛蟲公鑰$h(x)$和相關資訊(如權限(Permissions)) $I$ 得到密文 $c_1$，再把密文 $c_1$ 傳送給註冊中心。

當註冊中心收到密文 $c_1$ 後，用其私鑰解密得到終端設備的毛蟲公鑰$h(x)$和相關資訊 $I$。為了產製繭公鑰，將先產製另一個隨機多項式$d(x) \in R$，並且與毛蟲公鑰$h(x)$相乘後得到多項式作為繭公鑰$u(x)$，如公式(13)所示。之後，註冊中心再運用憑證中心公鑰加密繭公鑰$u(x)$和相關資訊 $I$ 得到密文 $c_2$，再把密文 $c_2$ 傳送給憑證中心。

$u(x) \in R_q \equiv h(x)d(x) \pmod{q}\pmod{x^N - 1}$ (13)

當憑證中心收到密文 $c_2$ 後，用其私鑰解密得到繭公鑰$u(x)$和相關資訊 $I$。為了產製蝴蝶公鑰，將先產製另一個隨機多項式$r(x) \in R$，並且與繭公鑰$u(x)$相乘後得到多項式作為蝴蝶公鑰$v(x)$，如公式(14)所示。之後，憑證中心產製終端設備的匿名憑證 $Cert_{EE}$，並且在匿名憑證 $Cert_{EE}$ 中的公鑰欄位存放蝴蝶公鑰$v(x)$，然後再用蝴蝶公鑰$v(x)$

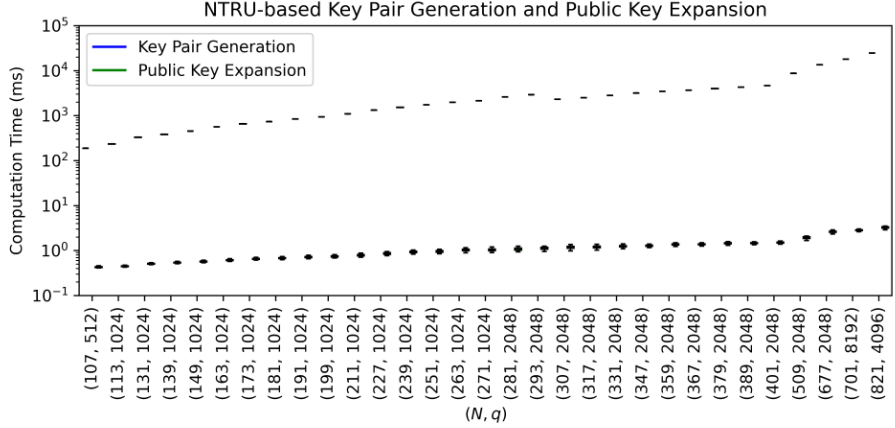

**Fig. 3.** The computation time of NTRU-based key pair generation and public key expansion (Unit: milliseconds).

加密匿名憑證 $Cert_{EE}$ 後得到密文 $c_3$，並經由註冊中心轉送密文 $c_3$ 給終端設備。

$$v(x) \in R_q \equiv u(x)r(x) \pmod{q}\pmod{x^N - 1} \quad (14)$$

為說明運用蝴蝶公鑰 $v(x)$ 加密匿名憑證 $Cert_{EE}$ 的過程中，本節假設匿名憑證 $Cert_{EE}$ 作為明文解碼為多項式的係數值產製待加密訊息多項式 $M(x) \in R_p$。透過公式(15)，產製隨機多項式 $b(x) \in R_q$，然後運用蝴蝶公鑰 $v(x)$ 對待加密訊息多項式 $M(x)$ 進行加密，產製多項式 $c_3(x)$ 作為密文。後續主要傳送多項式 $c_3(x)$ 的係數值。

$$\begin{aligned}c_3(x) &\in R_q \\ &\equiv p \times v(x) \times b(x) + M(x) \pmod{q}\pmod{x^N - 1} \\ &\equiv p \times F_q(x)g(x)d(x)r(x) \times b(x) + M(x) \\ &\qquad \pmod{q}\pmod{x^N - 1}\end{aligned} \quad (15)$$

當終端設備收到密文 $c_3$ 後，可以運用私鑰 $\{f(x), F_p(x)\}$ 解密得到匿名憑證 $Cert_{EE}$，計算過程如公式(16)、公式(17)、公式(18)。

$$\begin{aligned}c_3(x) \times f(x) &\in R_q \\ &\equiv p \times F_q(x)g(x)d(x)r(x) \times b(x) \times f(x) + M(x) \times f(x) \\ &\qquad \pmod{q}\pmod{x^N - 1} \\ &\equiv p \times g(x)d(x)r(x) \times b(x) + M(x) \times f(x) \\ &\qquad \pmod{q}\pmod{x^N - 1}\end{aligned} \quad (16)$$

$$\tau(x)'' \in R_p = c_3(x) \times f(x) \pmod{p}\pmod{x^N - 1} \quad (17)$$

$$\begin{aligned}\tau(x)'' \times F_p(x) &\in R_p \\ &\equiv p \times g(x)d(x)r(x) \times b(x) \times F_p(x) + M(x) \times f(x) \times F_p(x) \\ &\qquad \pmod{p}\pmod{x^N - 1} \\ &\equiv M(x) \in R_p\end{aligned} \quad (18)$$

2) 安全性分析

基於 NTRU 的匿名憑證方案的核心精神主要建構在繭公鑰(如公式(13)所示)和蝴蝶公鑰(如公式(14)所示)。由於從繭公鑰 $u(x)$ 反推毛蟲公鑰 $h(x)$ 和從蝴蝶公鑰 $v(x)$ 反推繭公鑰 $u(x)$ 可以歸納為最小向量問題[14]，並且目前尚無方法可以在多項式時間解決最小向量問題，從而保障攻擊者無法從繭公鑰 $u(x)$ 反推毛蟲公鑰 $h(x)$，也無法從蝴蝶公鑰 $v(x)$ 反推繭公鑰 $u(x)$，達到匿名性。

站在註冊中心的觀點，雖然註冊中心可以知道繭公鑰 $u(x)$ 和毛蟲公鑰 $h(x)$ 之間的關聯，但註冊中心無法從蝴蝶公鑰 $v(x)$ 反推繭公鑰 $u(x)$，所以註冊中心無法從蝴蝶公鑰 $v(x)$ 反推毛蟲公鑰 $h(x)$，故可以達到對註冊中心匿名。

站在憑證中心的觀點，雖然憑證中心可以知道蝴蝶公鑰 $v(x)$ 和繭公鑰 $u(x)$ 之間的關聯，但憑證中心無法從繭公鑰 $u(x)$ 反推毛蟲公鑰 $h(x)$，所以憑證中心無法從蝴蝶公鑰 $v(x)$ 反推毛蟲公鑰 $h(x)$，故可以達到對憑證中心匿名。

站在其他終端設備的觀點，無法從繭公鑰 $u(x)$ 反推毛蟲公鑰 $h(x)$，也無法從蝴蝶公鑰 $v(x)$ 反推繭公鑰 $u(x)$，故可以達到其他終端設備匿名。

綜合上述，本研究提出的基於 NTRU 的匿名憑證方案可以在最小向量問題的基礎上做到對全部設備匿名，保護終端設備的隱私。

## IV. 實驗結果與討論

為驗證提出的金鑰擴展方法，本研究採用硬體 RaspBerry Pi 4、函式庫 SageMath 和 Python 進行實作。本研究在 NTRU 密碼學方法採用的參數值為 $p = 3$，而 $N$ 和 $q$ 值以 "$(N, q)$" 來表示，圖 3 為不同參數組合下的金鑰對產製時間和金鑰擴展時間的結果，可以觀察到本研究所提方法可以提升約 436 倍~6553 倍的計算效率。

除此之外，本研究所提方法也跟 IEEE 1609.2.1 (即 ECC-based Method)[10]、Code-based BKE Method (即 McEliece-based Method)[15]，比較結果如表 I 所示。可以觀察到本研究所提方法可以具有較高的金鑰擴展效率，並且可以同時兼具量子安全。

## V. 結論與未來研究方向

本研究提出基於 NTRU 的後量子密碼學金鑰擴展方法和匿名憑證方案，可以提供高效率公鑰擴展，並且達到匿名性。實驗結果顯示計算效率可以高於現有已知方法，並且達到量子安全。

在未來研究可以嘗試在基於模格金鑰封裝機制和基於模格數位簽章演算法設計匿名憑證方案，以接軌標準。

TABLE I
COMPUTATION PERFORMANCE COMPARISON OF KEY EXPANSION METHODS (UNIT: MILLISECONDS)

| Method | Parameter Set | Security Level | Key Pair Generation | Public Key Expansion |
|---|---|---|---|---|
| ECC-based Method [10] | NIST P-256 | Unsecured | 23.308 | 23.319 |
| | NIST P-384 | Unsecured | 57.624 | 57.644 |
| | NIST P-521 | Unsecured | 115.043 | 114.788 |
| McEliece-based Method [15] | (12, 3884, 64) | 1 | 11752.678 | 1480.274 |
| | (13, 4608, 96) | 3 | 70714.403 | 7562.663 |
| | (13, 8192, 128) | 5 | 85677.689 | 11521.264 |
| The Proposed Method | (509, 2048) | 1 | 8721.401 | **1.929** |
| | (677, 2048) | 3 | 13529.710 | **2.597** |
| | (821, 4096) | 5 | 24558.711 | **3.748** |